\documentclass[letterpaper,preprintnumbers,english,floats,floatfix,amssymb,prd,superscriptaddress,twocolumn,nofootinbib]{revtex4}

\usepackage{amssymb,amsmath}
\usepackage{epsfig,hyperref}
\usepackage{epstopdf}
\usepackage[british]{babel}
\usepackage{enumerate}
\usepackage{enumitem}
\usepackage{color}

\def\be{\begin{equation}}
\def\ee{\end{equation}}
\def\beq{\begin{eqnarray}}
\def\eeq{\end{eqnarray}}

\makeatletter

\newcommand{\arXiv}[2][]{\href{http://arxiv.org/abs/#2}{\texttt{arXiv:#2\@ifempty{#1}{}{ [#1]}}}}
\makeatother

\begin{document}
\title{Energy in critical collapse}

\author{Yu Hu}%
\email{yuhu@hust.edu.cn}
\affiliation{MOE Key Laboratory of Fundamental Physical Quantities Measurement, Hubei Key Laboratory of Gravitation and Quantum Physics, PGMF, and School of Physics, Huazhong University of Science and Technology, Wuhan 430074, Hubei, China}

\author{Jun-Qi Guo}%
\email{sps{\_}guojq@ujn.edu.cn}
\affiliation{School of Physics and Technology, University of Jinan, Jinan 250022, Shandong, China}

\author{Junbin Li}%
\email{lijunbin@mail.sysu.edu.cn}
\affiliation{Department of Mathematics, Sun Yat-sen University, Guangzhou 510275, Guangdong, China}

\author{Cheng-Gang Shao}%
\email{cgshao@hust.edu.cn}
\affiliation{MOE Key Laboratory of Fundamental Physical Quantities Measurement, Hubei Key Laboratory of Gravitation and Quantum Physics, PGMF, and School of Physics, Huazhong University of Science and Technology, Wuhan 430074, Hubei, China}

\author{Hongsheng Zhang}%
\email{sps{\_}zhanghs@ujn.edu.cn}
\affiliation{School of Physics and Technology, University of Jinan, Jinan 250022, Shandong, China}

\date{\today}

\begin{abstract}
  We study the energy issue in critical collapse of a spherically symmetric scalar field. It is found that in critical collapse, the contribution from the material energy is greater than that from the gravitational energy. The quantity $m/r$ plays an important role in identifying the formation of apparent horizon in gravitational collapse, where $m$ is the Misner-Sharp mass and $r$ the areal radius. We observe that in critical collapse, the maximum value of $m/r$ fluctuates between $2/15$ and $4/15$. This denotes a large gap between critical collapse and black hole formation for which the criterion is $m/r=1/2$.
  \end{abstract}
\maketitle

\section{Introduction}
Energy is a key subject in physics. Due to the equivalence principle, one cannot define the gravitational energy locally. The success of the proof for the positive energy theorem inspired people to define the gravitational energy at the quasi-local level: the energy contained in a closed two-dimensional surface. Several notions of the quasi-local energy have been constructed, such as the Misner-Sharp energy~\cite{Misner_1964}, Brown-York energy~\cite{Brown:1992br}, Hawking energy~\cite{Hawking_1968}, Hayward energy~\cite{Hayward:1993ph}, and Wang-Yau energy~\cite{Wang:2008jy,Wang:2009}. The descriptions for the energy of gravitational waves were studied in Refs.~\cite{PhysRev.166.1263,PhysRev.166.1272,Cai:2021jbi}. More details can be found in review~\cite{Szabados_2009}.

By fine-tuning the initial data of a spherically scalar field, Choptuik discovered the critical phenomena in gravitational collapse~\cite{Choptuik_1993}. On the sub-critical side, the collapse will lead to dispersion, while on the super-critical side, a tiny black hole will form. The critical collapse solution is discretely self-similar along the ever-decreasing time and space scales. This feature is universal, independent of the initial profiles of the matter field. The mass of the tiny black hole forming in the super-critical circumstance satisfies a power law, $M_{BH}\propto|p-p^*|^\gamma$, where $p^*$ is a critical parameter value in the initial data of the scalar field, and $\gamma\approx0.37$. Critical phenomena in many other collapse models were also observed (see \cite{Gundlach_2007} for review). Recently, the critical behavior in 3D gravitational collapse with no symmetry assumptions was studied in Ref.~\cite{Deppe:2018uye}. Critical collapse of electromagnetic waves in axisymmetry was investigated in Refs.~\cite{Baumgarte:2019fai,Mendoza:2021nwq}. Novel dynamical critical phenomena in the process of the nonlinear accretion of the scalar field into black holes were observed in Ref.~\cite{Zhang:2021nnn}. With one typical log-periodic formula in the discrete scale invariance systems, one approximate analytic solution for the spacetime near the center was obtained in Ref.~\cite{Guo:2018yyt}.

Three versions of the quasi-local energy in the Oppenheimer-Snyder dust collapse model were analyzed in Ref.~\cite{He:2020ygx}. Quasi-local energy was calculated in the contexts of black hole physics and cosmology in Refs.~\cite{Afshar:2009pi,Miller:2017udt,Gudapati:2020cqb}. In Ref.~\cite{Liang}, the gravitational and material energies in a static spherical star were discussed. Considering the fundamental role that energy has been playing in physics, in this paper, we investigate the energy issue in critical collapse, and we are especially interested in the comparison between the contributions from the gravitational and material energies.

This paper is organized as follows. In Sec.~\ref{sec:methodology}, we describe the methodology, including the collapse model and the definitions of energy. In Sec.~\ref{sec:energy}, we discuss the energy issue in critical collapse. Some features of the quantity $m/r$ are studied in Sec.~\ref{sec:m_r}. In Sec.~\ref{sec:summary}, the results are summarized.

\section{Methodology\label{sec:methodology}}
\subsection{Gravitational collapse}
We consider critical collapse of a spherically symmetric massless scalar field $\phi$. The action for system is
\be S=\int \mbox{d}^{4}x\sqrt{-g}\left(\frac{R}{16\pi G}-\frac{1}{2}\nabla^{\mu}\phi\nabla_{\mu}\phi\right). \label{action_flat}\ee
We set $G=1$. The corresponding energy-momentum tensor for $\phi$ is
\be T_{\mu\nu}=\phi_{,\mu}\phi_{,\nu}-\frac{1}{2}g_{\mu\nu}g^{\alpha\beta}\phi_{,\alpha}\phi_{,\beta}.\label{tensor_phi}\ee
We simulate the critical collapse in polar coordinates,
\be ds^{2}=-A(r,t)\mbox{e}^{-2\delta(r,t)}\mbox{d}t^{2}+\frac{1}{A(r,t)}\mbox{d}r^{2}+r^{2}\mbox{d}\Omega^2.\label{metric}\ee

For the metric~(\ref{metric}), some components of the Einstein tensor and the energy-momentum tensor for $\phi$ are shown below:
\be G^{t}_{t}=\frac{1}{r^2}(rA_{,r}-1+A),\nonumber\ee
\be G^{r}_{r}=-\frac{1}{r^2}(-rA_{,r}+2rA\delta_{,r}+1-A),\nonumber\ee
\be G^{r}_{t}=-\frac{1}{r}A_{,t},\nonumber\ee
\be T^{t}_{t}=-T^{r}_{r}=-\frac{1}{2}A(P^2+Q^2),\nonumber\ee
\be T^{r}_{t}=A\phi_{,r}\phi_{,t}.\nonumber\ee
Here, $(_{,r})$ and $(_{,t})$ denote partial derivatives with respect to the coordinates $r$ and $t$, respectively.

We define
\be Q(r,t)\equiv\phi_{,r}, \hphantom{dd} P(r,t)\equiv A^{-1}\mbox{e}^{\delta}\phi_{,t}.\label{define_Q_P}\ee
Then, the equation of $G^{t}_{t}=8\pi T^{t}_{t}$ yields
\be A_{,r}=\frac{1-A}{r}-4\pi rA(P^{2}+Q^{2}).\label{eomA}\ee
Combining $G^{r}_{r}=8\pi T^{r}_{r}$ and Eq.~(\ref{eomA}), we have
\be \delta_{,r}=-4\pi r(P^{2}+Q^{2}).\label{eomd}\ee
The equation of $G^{r}_{t}=8\pi T^{r}_{t}$ generates 
\be A_{,t}=-8 \pi rA^{2}\mbox{e}^{-\delta}PQ.\label{eqAt}\ee
With Eq.~(\ref{define_Q_P}), we obtain
\be Q_{,t}=(A \mbox{e}^{-\delta} P)_{,r}.\label{eomQ}\ee
For the metric~(\ref{metric}), the conservation of the energy-momentum tensor, $T^{\mu\nu}_{;\mu}=0$, leads to
\be P_{,t}=\frac{1}{r^{2}}(r^{2} A \mbox{e}^{-\delta} Q)_{,r}.\label{eomP}\ee

The initial conditions for $\phi$ are set as $\phi|_{t=0}=a\exp[-(r/\sigma)^{2}]$, and $\phi_{,t}|_{t=0}=0$. The spacial range is $0\le r\le 12$. In seeking the numerical solution to critical collapse, we set $a=0.336033778324$ and $\sigma=1$. The initial values for the metric functions $A$ and $\delta$ are obtained via integrations of Eqs.~\eqref{eomA} and \eqref{eomd}. The regularity of Eq.~\eqref{eomA} at the center requires that $A|_{r=0}=1$. We choose $\delta|_{r=0}=0$. Consequently, the coordinate time is equal to the proper time at the center.

In the simulation, we integrate Eqs.~(\ref{eomA}), (\ref{eomd}), (\ref{eomQ}) and (\ref{eomP}) using the fourth-order Runge-Kutta method. A mesh refinement algorithm is implemented. For details on the numerics, see Ref.~\cite{Zhang:2016kzg}.

\subsection{Landau-Lifshitz pseudotensor approach}
We implement two separate approaches to define the gravitational and material energies: Landau-Lifshitz pseudotensor approach and Misner-Sharp energy approach. The Landau-Lifshitz pseudotensor is a typical definition for the energy of the gravitational field, and is coordinate-dependent. The Misner-Sharp energy is one of the major notions for the quasi-local energy enclosed in a two-dimensional spacelike surface in spherical symmetry, and is coordinate-independent.

Regarding the first approach, we firstly take the following definitions~\cite{Landau_Lifshitz,Poisson},
\be \mathfrak{g}^{\alpha\beta}\equiv\sqrt{-g}g^{\alpha\beta},\ee
\be H^{\mu\alpha\nu\beta}\equiv\mathfrak{g}^{\mu\nu}\mathfrak{g}^{\alpha\beta}-\mathfrak{g}^{\mu\beta}\mathfrak{g}^{\alpha\nu},\ee
\be
\begin{split}
&(-g)t_{LL}^{\alpha\beta}\equiv\\
&\frac{1}{16\pi}
\Big\{\partial_{\lambda}\mathfrak{g}^{\alpha\beta}\partial_{\mu}\mathfrak{g}^{\lambda\mu}
-\partial_{\lambda}\mathfrak{g}^{\alpha\lambda}\partial_{\mu}\mathfrak{g}^{\beta\mu} \\
&+\frac{1}{2}g^{\alpha\beta}g_{\lambda\mu}\partial_{\rho}\mathfrak{g}^{\lambda\nu}\partial_{\nu}\mathfrak{g}^{\mu\rho}
-g^{\alpha\lambda}g_{\mu\nu}\partial_{\rho}\mathfrak{g}^{\beta\nu}\partial_{\lambda}\mathfrak{g}^{\mu\rho}\\
&-g^{\beta\lambda}g_{\mu\nu}\partial_{\rho}\mathfrak{g}^{\alpha\nu}\partial_{\lambda}\mathfrak{g}^{\mu\rho}
+g_{\lambda\mu}g^{\nu\rho}\partial_{\nu}\mathfrak{g}^{\alpha\lambda}\partial_{\rho}\mathfrak{g}^{\beta\mu}\\
&+\frac{1}{8}(2g^{\alpha\lambda}g^{\beta\mu}-g^{\alpha\beta}g^{\lambda\mu})(2g_{\nu\rho}g_{\sigma\tau}-g_{\rho\sigma}g_{\nu\tau})
\partial_{\lambda}\mathfrak{g}^{\nu\tau}\partial_{\mu}\mathfrak{g}^{\rho\sigma}
\Big\},
\end{split}\label{tll}
\ee
where $g$ is the metric determinant, and $t_{LL}^{\alpha\beta}$ is called the Landau-Lifshitz pseudotensor. The tensor density $H^{\alpha\mu\beta\nu}$ satisfies the identity
\be
\partial_{\mu\nu}H^{\alpha\mu\beta\nu}=2(-g)G^{\alpha\beta}+16\pi(-g)t_{LL}^{\alpha\beta}.
\ee
Then the Einstein equations can be expressed in the non-tensorial form
\be \partial_{\mu\nu}H^{\alpha\mu\beta\nu}=16\pi(-g)(T^{\alpha\beta}+t_{LL}^{\alpha\beta}).\label{EFE_H}\ee
Using the antisymmetric property of $H^{\mu\alpha\nu\beta}$ and the Einstein field equations, one obtains
\be \partial_{\beta}\left[(-g)(T^{\alpha\beta}+t_{LL}^{\alpha\beta})\right]=0.\label{conserve}\ee
Equations~(\ref{EFE_H}) and (\ref{conserve}) imply that $t_{LL}^{\alpha\beta}$ can be interpreted as an energy-momentum (pseudo)tensor for the gravitational field.

The quantity $t_{LL}^{\alpha\beta}$ is nontensorial, and is usually considered to be meaningless. However, the nontensorial nature of an object does not imply that it is meaningless. The Christoffel symboles are nontensorial. However, they do have geometric and physical content. Although these coefficients can be taken to be zero at a given point by coordinate transformation, they cannot be transformed to zero on an open domain in curved spacetime~\cite{Szabados_2009}. Consequently, we think that $t_{LL}^{\alpha\beta}$ may carry some useful information.

We define a total energy associated with the region $V$,
\be E_{LL}(V)\equiv\int_{V}(-g)(T^{00}+t_{LL}^{00})\mbox{d}^{3}x.
\label{LLV}
\ee
When $V$ includes the whole space, $E_{LL}(V)$ coincides with the Arnowitt-Deser-Misner mass.
Since the value of the pseudotensor is coordinate-dependent, we choose the coordinates carefully when using a pseudotensor. Here we prefer the Cartesian coordinates, in which
for the metric \eqref{metric}, there are
\be
g_{tt}=-A \mbox{e}^{-2\delta}, \hphantom{ddd}g_{ij}=\delta_{ij}+\frac{x_{i}x_{j}(1-A)}{r^{2}A}.
\label{metricCar}
\ee
Substitution of Eq.~\eqref{metricCar} into \eqref{LLV} yields
\be
E_{LL}(V)=\frac{r}{2}\left(\frac{1}{A}-1\right).
\label{LL}
\ee
Substituting Eq.~(\ref{metricCar}) into (\ref{tll}) and using $g\equiv|g_{\mu \nu}|=-\mbox{e}^{-2\delta}$, we obtain
\be (-g)t_{LL}^{00}=\frac{(A-1)A_{,r}}{8\pi rA^{2}}.\ee
With Eq.~(\ref{tensor_phi}), we have $T^{00}=(1/2)\mbox{e}^{-2\delta}(P^{2}+Q^{2})$. Then we can split the total energy (\ref{LLV}) into the material and gravitational parts,
\be
E_{LL}^{m}(V)\equiv\int_{V}(-g)T^{00}\mbox{d}^{3}x=4\pi\int^{r}_{0}\frac{1}{2}(P^{2}+Q^{2})r^{2}\mbox{d}r,
\label{ELLm}
\ee
\be
E_{LL}^{g}(V)\equiv\int_{V}(-g)t_{LL}^{00}\mbox{d}^{3}x=4\pi\int^{r}_{0}\frac{(A-1)A_{,r}}{8\pi rA^{2}}r^{2}\mbox{d}r.
\label{ELLg}
\ee

\subsection{Misner-Sharp energy approach}
Before discussing the definitions for the material and gravitational energies in critical collapse with the Misner-Sharp energy approach, we firstly consider the energy issue in a static star of perfect fluid, with the energy-momentum tensor, $T_{\mu\nu}=(\rho_{f}+p_{f})U_{\mu}U_{\nu}+p_{f}g_{\mu\nu}$, where $U^{\mu}$ is the tangent vector of the stationary observer. Inside the star, one mass function can be defined as
\be m(r)\equiv4\pi\int^{r}_{0}\rho_{f}r'^2\mbox{d}r'=\frac{r}{2}(1-g_{rr}^{-1}).\ee
On the boundary of the star $r=r_b$, there is
\be M=m(r_b)=4\pi\int^{r_b}_{0}\rho_{f}r'^{2}\mbox{d}r',\label{M_star}\ee
which is identical to the expression for the total mass in Newtonian gravity. However, in general relativity, the proper mass is
\be
M_p=\int\rho_{f}\sqrt{h}\mbox{d}r'\wedge \mbox{d}\theta\wedge \mbox{d}\phi=4\pi\int^{r_b}_{0}\rho_{f}A^{-1/2}r'^{2}\mbox{d}r',
\label{M_proper_star}\ee
where $h(=A^{-1}r'^4\sin^{2}\theta)$ is the determinant of the induced metric $h_{ab}[=\mbox{diag}(A^{-1},r'^2,r'^2\sin^{2}\theta)]$ of the spacelike hypersurface $t=\text{Const}$.

The quantity $M$ in Eq.~(\ref{M_star}) has clear physical meaning: it is the total mass (energy) of the Schwarzschild spacetime, including gravitational potential energy. However, the quantity $M_p$ in Eq.~(\ref{M_proper_star}) does not include gravitational potential energy, since $\rho_f$ (the energy density measured by a stationary observer) includes the rest energy density of each particle in the star and the internal energy density, but does not include the gravitational energy density~\cite{Liang}.

Regarding the Misner-Sharp energy approach of defining the material and gravitational energies in critical collapse, we take similar method as in the static circumstance  discussed above. We use the Misner-Sharp energy~\cite{Misner_1964} to represent the total energy inside a closed two-dimensional spacelike surface $S$ constrained by $r=\text{Const}$ and $t=\text{Const}$.
\be
\begin{split}
E^{\text{total}}_{MS}(r)&\equiv\frac{r}{2}(1-g^{\mu\nu}r_{,\mu}r_{,\nu})\\
&=\frac{r}{2}(1-A)\\
&=4\pi\int^{r}_{0}\rho r'^{2}\mbox{d}r'\\
&=4\pi\int^{r}_{0}\frac{1}{2}A(P^{2}+Q^{2})r'^{2}\mbox{d}r',
\end{split}
\label{MS}
\ee
where $\rho$ is the material energy density measured by a stationary observer, $\rho\equiv T_{\mu\nu}U^{\mu}U^{\nu}=(1/2)A(P^{2}+Q^{2})$. $U^{\mu}$ is the tangent vector of the stationary observer, $U^{\mu}=(\sqrt{-g^{tt}},0,0,0)$. The material and gravitational energies are respectively defined as below,
\be
E^{m}\equiv\int\rho\sqrt{h}\mbox{d}r'\wedge \mbox{d}\theta\wedge \mbox{d}\phi=4\pi\int^{r}_{0}\frac{1}{2}\sqrt{A}(P^{2}+Q^{2})r'^{2}\mbox{d}r',
\label{Em}
\ee
\be
E^{g}_{MS}\equiv E^{\text{total}}_{MS}-E^{m}=4\pi\int^{r}_{0}\frac{1}{2}(A-\sqrt{A})(P^{2}+Q^{2})r'^{2}\mbox{d}r'.
\label{Eg}
\ee

\begin{figure}[!htbp]
    \centering
    \includegraphics[width=0.36\textwidth]{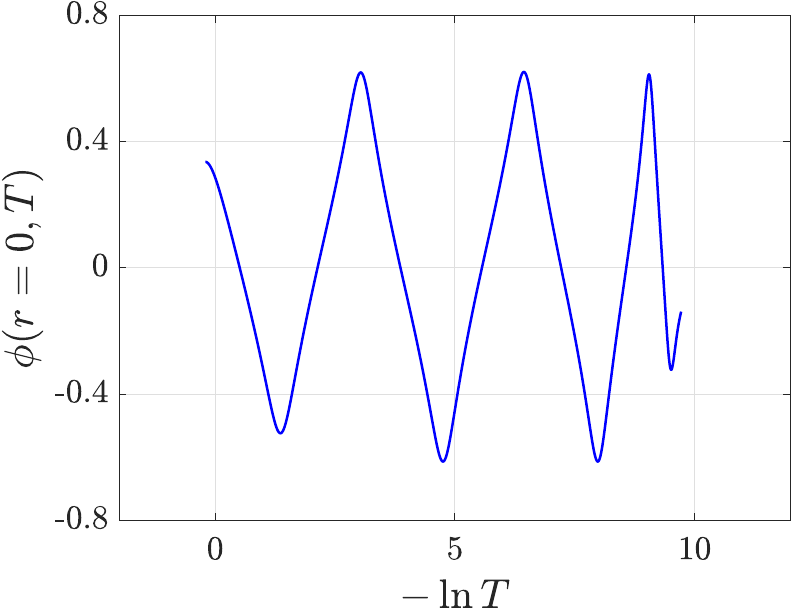}
    \caption{Evolution of the scalar field at the center in critical collapse.}
    \label{fig:scalar}
\end{figure}

\begin{figure*}[!htbp]
    \centering
    \includegraphics[width=0.82\textwidth]{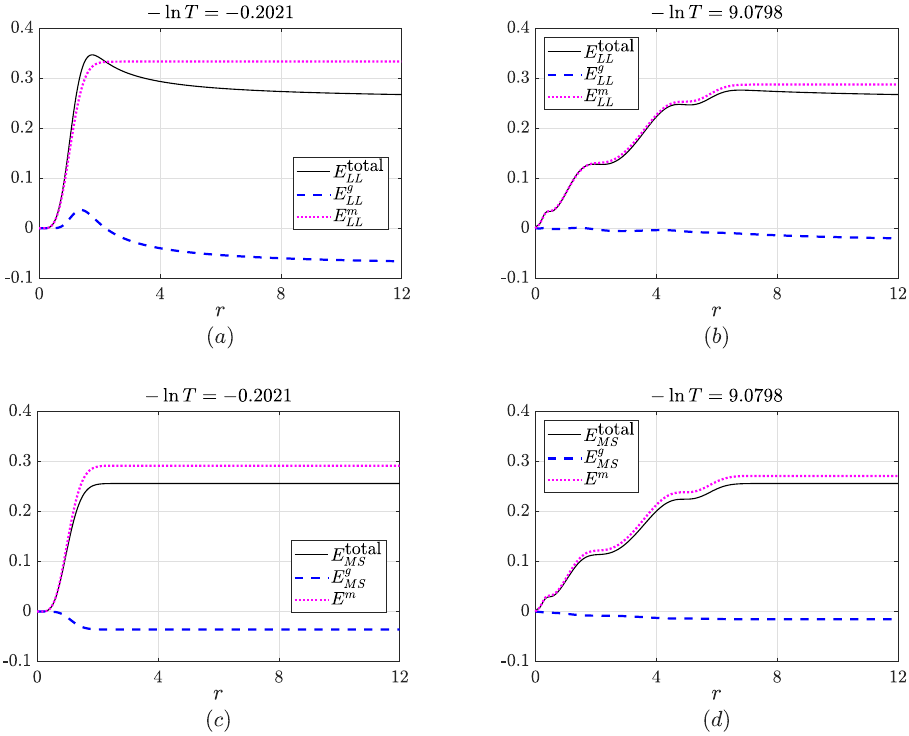}
    \caption{(color online) Gravitational and material energies in critical collapse defined by Eqs.~(\ref{LLV}), (\ref{ELLm}), (\ref{ELLg}), (\ref{MS})-(\ref{Eg}).}
    \label{fig:energy_parts_critical}
\end{figure*}

\begin{figure*}[!htbp]
    \centering
    \includegraphics[width=0.84\textwidth]{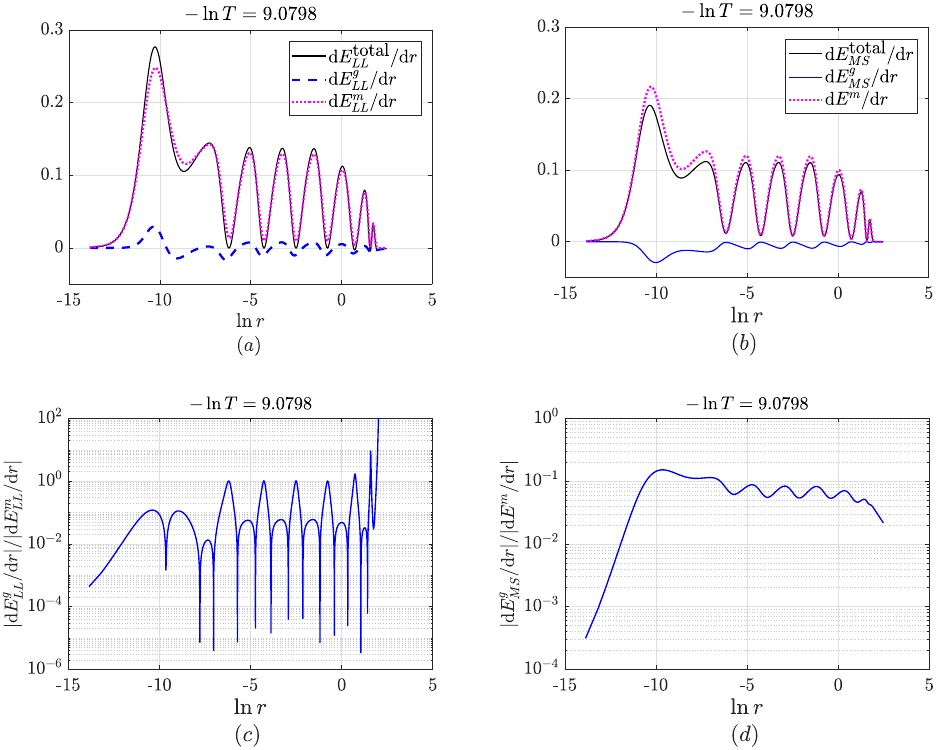}
    \caption{(color online) Energy density at the critical collapse stage. The absolute value of the ratio between the gravitational and material energy densities in the large-radius region is much greater than that in the small-radius one.}
    \label{fig:density}
\end{figure*}

\begin{figure*}[!htbp]
    \centering
    \includegraphics[width=0.82\textwidth]{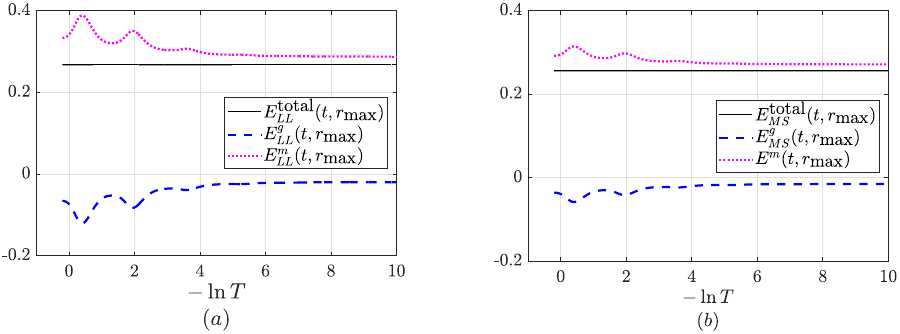}
    \caption{(color online) Evolution of the energy inside the region $r\le12$.}
    \label{fig:energy_evolution}
\end{figure*}

\begin{figure}[!htbp]
    \centering
    \includegraphics[width=0.39\textwidth]{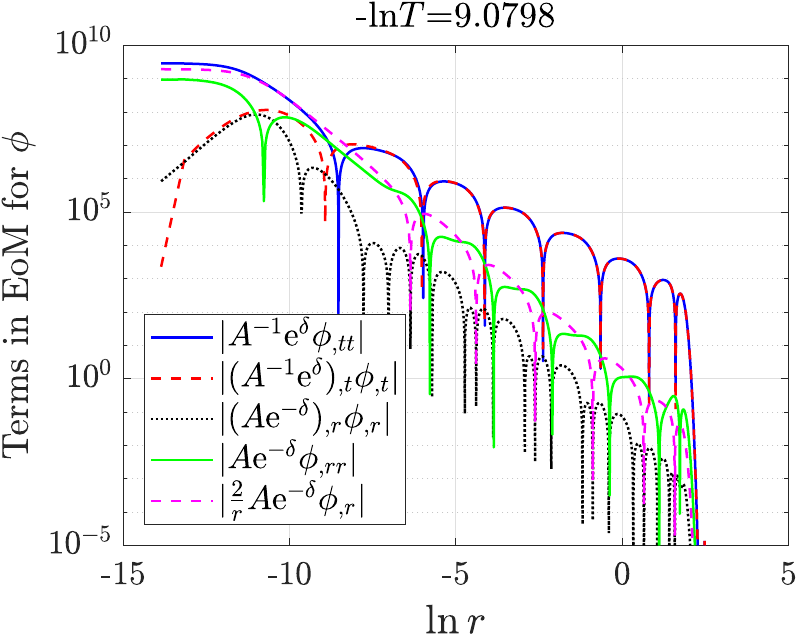}
    \caption{(color online) Terms in the equation of motion for $\phi$ (\ref{eomP}). In the small- and large-radius regions, Eq.~(\ref{eomP}) is respectively reduced to $\phi_{,tt}\approx r^{-2}(r^{2}\phi_{,r})_{,r}$ and $A^{-1}\mbox{e}^{\delta}\phi_{,tt}\approx -(A^{-1}\mbox{e}^{\delta})_{,t}\phi_{,t}$.}
    \label{fig:eom_phi}
\end{figure}

\begin{figure*}[!htbp]
    \centering
    \includegraphics[width=0.85\textwidth]{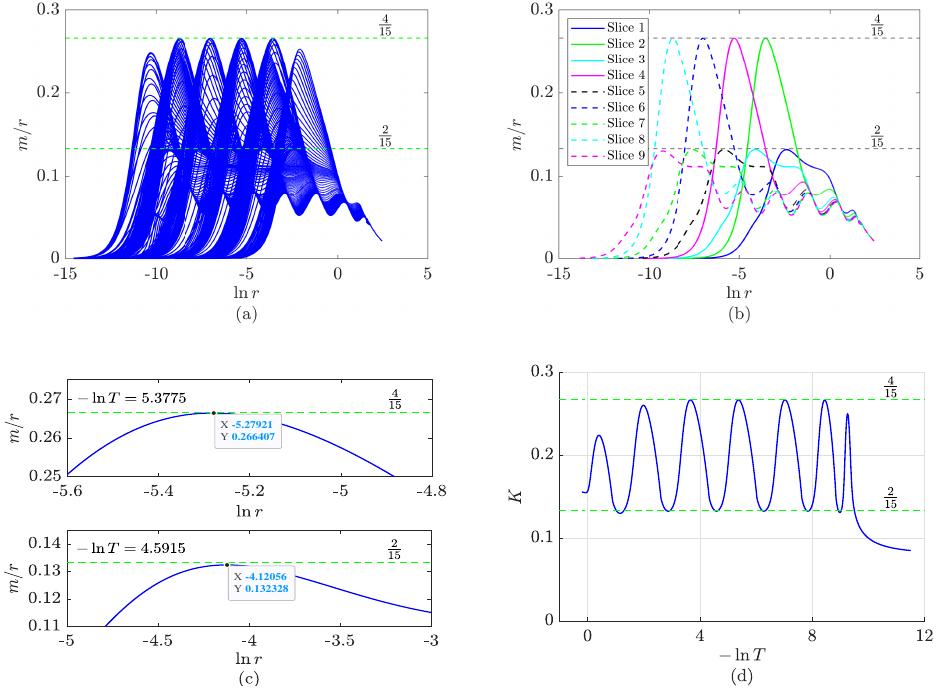}
    \caption{(color online) $m/r$ in critical collapse in the polar coordinates~(\ref{metric}). (a) and (b): $m/r$ on some slices $t=\text{Const}$. In the transition place between the small- and large-radius regions, the maximum values of $m/r$ stay in the range of $[2/15, 4/15]$. In Fig.~(b), the slices are numbered in the temporal direction.
    (c) Numerical illustrations on the upper and lower limits for the maximum values of $m/r$. (d) $K$: the maximum value of $m/r$ on slices $T=\text{Const}$.}
    \label{fig:m_r}
\end{figure*}

\begin{figure*}[!htbp]
    \centering
    \includegraphics[width=0.85\textwidth]{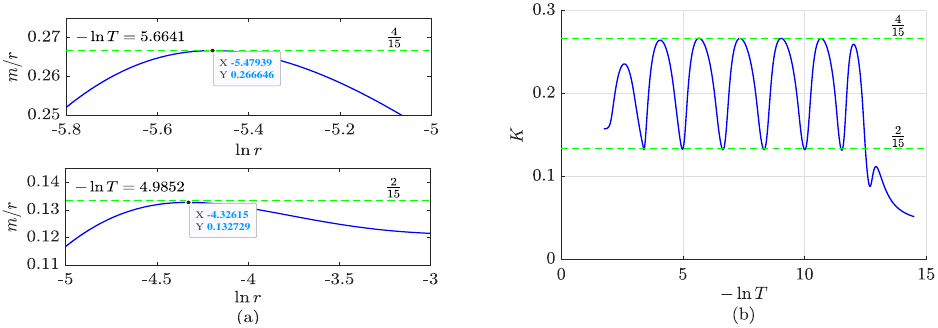}
    \caption{(color online) $m/r$ in critical collapse on some slices $t=\text{Const}$ in the double-null coordinates~(\ref{double_null_metric}). (a) Numerical illustrations on the upper and lower limits for the maximum values of $m/r$. (b) $K$: the maximum value of $m/r$ on slices $T=\text{Const}$. }
    \label{fig:m_r_null}
\end{figure*}

\section{Results I: energy\label{sec:energy}}
By fine-tuning the initial data of the scalar field, we obtain the numerical solution to critical collapse. The oscillating behavior of the scalar field at the center, $\phi$ vs. $-\ln T$, is shown in Fig.~\ref{fig:scalar}, where $T\equiv t_{*}-t$, and $t_{*}$ is the time for naked singularity formation. The period $\Delta$ for the oscillations takes the same value as reported in the literature, $\Delta\approx 3.43$.

With the expressions (\ref{LLV}), (\ref{ELLm}), (\ref{ELLg}), (\ref{MS})-(\ref{Eg}), we plot the material and gravitational energies and energy densities in Figs.~\ref{fig:energy_parts_critical} and \ref{fig:density}, respectively. We also plot the evolution of the energies in a sufficiently large region of $r\le12$ in Fig.~\ref{fig:energy_evolution}, such that during the simulation the boundary is far enough from the collapse region and the total energy in the region remains constant. From these figures, we observe that both the Landau-Lifshitz pseudotensor approach and Misner-Sharp energy approach generate similar results:
\begin{enumerate}[fullwidth,itemindent=0em,label=(\roman*)]
  \item In critical collapse, the contribution from the material energy is greater than that from the gravitational energy.
  \item The material energy density, $\mbox{d}E^{m}_{LL}/\mbox{d}r$ in Fig.~\ref{fig:density}(a) and $\mbox{d}E^{m}/\mbox{d}r$ in Fig.~\ref{fig:density}(b), is positive.
  \item As shown in Fig.~\ref{fig:density}(a), the gravitational energy density $\mbox{d}E^{g}_{LL}/\mbox{d}r$ by the Landau-Lifshitz pseudotensor approach is sometimes negative and sometimes positive. As shown in Fig.~\ref{fig:density}(b), the gravitational energy density $\mbox{d}E^{g}_{MS}/\mbox{d}r$ by the Misner-Sharp energy approach is always negative.
\end{enumerate}

There is a strong correlation between the gravitational effects on the evolution of the scalar field and the ratio between the gravitational and material energy densities. As shown in Fig.~\ref{fig:eom_phi}, in the large-radius region ($\ln r>-7$), the equation of motion for $\phi$~(\ref{eomP}) is reduced to
\be A^{-1}\mbox{e}^{\delta}\phi_{,tt}\approx -(A^{-1}\mbox{e}^{\delta})_{,t}\phi_{,t}.\label{eomP_large}\ee
Therefore, the gravitational effects on the dynamics of the scalar field are important. On the other hand, in the small-radius region ($\ln r<-11$), Eq.~(\ref{eomP}) becomes
\be \phi_{,tt}\approx r^{-2}(r^{2}\phi_{,r})_{,r},\label{eomP_center}\ee
which is similar to what happens in flat spacetime. So the scalar field does not feel the gravitational effects directly. On the other hand, as shown in Fig.~\ref{fig:eom_phi}, in the transition region locating between the central and large-radius ones, in the equation of motion for the scalar field, the gravitational effects are not negligible. Then due to the connection between the central and transition regions, the scalar field in the central region feels the gravitational effects indirectly. Correspondingly, as shown in Fig.~\ref{fig:density}, the absolute value of the ratio between the gravitational and material energy densities in the large-radius region is much greater than in the small-radius one. The echoing behavior of the energy density is clearly demonstrated in Fig.~\ref{fig:density}.

We make some comments on the result~(\ref{eomP_center}). In another two types of collapse (dispersion and early stage of collapse toward black hole formation), near the center, Eq.~(\ref{eomP}) is also reduced to the form (\ref{eomP_center}). The causes are the following. In Eq.~(\ref{eomP}), the gravitational effects come from the first-order derivatives of the metric functions. Under the smoothness requirement in the central region, the metric functions and the scalar field have the following asymptotic expressions~\cite{Zhang:2016kzg}:
\be A\approx 1+A_{2}(t)r^2, \hphantom{d}\delta\approx\delta_{2}(t)r^2, \hphantom{d}\phi\approx\phi_0 + \phi_{2}(t)r^2.\nonumber\ee
Some details on the analytic investigations of Eqs.~(\ref{eomP_large}) and (\ref{eomP_center}) are presented in Ref.~\cite{Guo:2023cya}.

As discussed in Ref.~\cite{Guo:2023cya}, in the large-radius region, the field $\phi$ admits the following approximate expression:
\be \phi(r,\xi)\approx C_{1}(1+C_{2}[H(r,\xi)])\cos(\omega\ln r + C_{3}[H(r,\xi)] + \varphi_{0} ),\label{phi_sln}\ee
where $\xi\equiv t-t_{*}$, such that $\xi=0$ upon naked singularity formation. The quantity $[H(r,\xi)]$ has the following features:
\begin{enumerate}[fullwidth,itemindent=0em,label=(\roman*)]
  \item For $[H(r,\xi)]$, there is
  \be [H(r,\xi)]=H(r,\xi)\equiv\frac{\omega\alpha \xi}{r}=\omega A^{1/2}e^{-\delta}\frac{\xi}{r},\ee
  where $\alpha\equiv A^{1/2}e^{-\delta}$.
  \item Note that $H_{,\xi}=\omega\alpha/r+\omega\alpha_{,\xi}\xi/r$. The numerical results yield
  \be [H]_{,\xi}\approx\frac{\omega\alpha}{r},\hphantom{dd} [H]_{,\xi\xi}\approx\frac{\omega\alpha_{,\xi}}{r}\approx -\delta_{,\xi}[H]_{,\xi}.\label{H_tt}\ee
\end{enumerate}

The numerical results show that the transition region locating between the central and large-radius regions can be described as $r\in [r_1,~r_2]$. At $r=r_{1}$, there is
$|C_{3}H|\sim|\omega\ln r|$; and at $r=r_2$, there is $|C_{3}H_{,r}|\sim \omega/r$.

\section{Result II: $m/r$\label{sec:m_r}}
In gravitational collapse, $m/r$ is an important quantity identifying the location of apparent horizon, which is also crucial in identifying the formation of singularities~\cite{Chris_1993}, where $m$ is the Misner-Sharp mass. It is natural to ask how far the spacetime in critical collapse is from black hole formation. Moreover, as studied in Ref.~\cite{Price:1996sk}, the quantity $m/r$ is closely related to the origin of the characteristic period in critical collapse $\Delta\approx 3.43$.

We plot the evolution of $m/r$ on some slices $t=\mbox{Const}$ in Fig.~\ref{fig:m_r}. On each slice $t=\mbox{Const}$, the quantity $m/r$ takes its maximum value, denoted as $(m/r)|_{\scriptsize\mbox{max}}$, in the transition place between the small- and large-radius regions. In Ref.~\cite{Choptuik_workshop_1993}, it was reported that the maximum value of $(m/r)|_{\scriptsize\mbox{max}}$ is $0.26$. In this work, we obtain more accurate results. Denote $K(T)$ as the maximum value of $m/r$ on slices $T=\text{Const}$. As shown in Fig.~\ref{fig:m_r}, the numerical results for the upper and lower limits for $K$ are about $0.266407$ and $0.132328$, respectively. With the same code used in Ref.~\cite{Guo:2020ked}, we simulate critical collapse in the double-null coordinates,
\be
\begin{split}
ds^{2} &= -4e^{-2\sigma(u,v)}\mbox{d}u\mbox{d}v+{r^{2}(u,v)}\mbox{d}\Omega^2\\
&= \mbox{e}^{-2\sigma(t,x)}(-\mbox{d}t^2+\mbox{d}x^2)+{r^{2}(t,x)}\mbox{d}\Omega^2,
\end{split}
\label{double_null_metric}
\ee
where $u=(t-x)/2$ and $v=(t+x)/2$. As shown in Fig.~\ref{fig:m_r_null}, we obtain the numerical results for the upper and lower limits for $K$ which are $0.266646$ and $0.132729$, respectively. So it is natural to state the limits by the fractional numbers, $4/15$ and $2/15$.

There is a big gap between the upper limit for $m/r$ in critical collapse (which is $4/15$) and the criterion for black hole formation (which is $m/r=1/2$). This is reasonable since dispersion, critical collapse and black hole formation are three distinct final outcomes of gravitational collapse.

\section{Summary\label{sec:summary}}
Energy has been playing a fundamental role in physics, and the explorations on this concept keep bringing us insightful understanding on Nature. However, in general relativity, because of the nontensorial characteristic of the gravitational energy-momentum density expression, the energy issue in gravitational collapse has not been fully studied in the literature. A nontensorial object may still be meaningful. The Christoffel symboles are nontensorial and one can make them be zero at a given point by coordinate transformation. However, they cannot be transformed to zero on an open domain in curved spacetime. So in this paper we took the adventure of studying the energy issue in critical collapse with the Landau-Lifshitz pseudotensor approach and Misner-Sharp energy approach. These two approaches generate similar results: in critical collapse, the contribution from the material energy is greater than that from the gravitational energy.

The quantity $m/r$ is indispensable in identifying the formation of apparent horizon in gravitational collapse. In this paper, it was observed that in critical collapse the maximum value of $m/r$ fluctuates between $2/15$ and $4/15$. So the upper bound $4/15$ for $m/r$ is a bit far from the criterion for black hole formation, $m/r=1/2$.

\section*{Acknowledgments}\small
The authors are very grateful to the anonymous referees for the valuable comments. The authors thank Xiaokai He, Xiaoning Wu, and Cheng-Yong Zhang for the helpful discussions. YH and CGS are supported by the National Natural Science Foundation of China (Grant No. 11925503). JQG is supported by Shandong Province Natural Science Foundation under grant No.ZR2019MA068.


\begin{thebibliography}{99}

\bibitem{Misner_1964}
C. W. Misner and D. H. Sharp,
Phys. Rev. {\bf 136}, B571 (1964).

\bibitem{Brown:1992br}
J. D. Brown and J. W. York, Jr.,
Phys. Rev. D {\bf 47}, 1407 (1993).
[\arXiv{gr-qc/9209012}]

\bibitem{Hawking_1968}
S. W. Hawking,
J. Math. Phys. {\bf 9}, 598 (1968).

\bibitem{Hayward:1993ph}
S. A. Hayward,
Phys. Rev. D {\bf 49}, 831 (1994).
[\arXiv{gr-qc/9303030}]

\bibitem{Wang:2008jy}
M.-T. Wang and S.-T. Yau,
Phys. Rev. Lett. {\bf 102}, 021101 (2009).
[\arXiv[gr-qc]{0804.1174}]

\bibitem{Wang:2009}
M.-T. Wang and S.-T. Yau,
Commun. Math. Phys. {\bf 288}, 919 (2009).
[\arXiv[math.DG]{0805.1370}]

\bibitem{PhysRev.166.1263}
R. A. Isaacson,
Phys. Rev. {\bf 166}, 1263 (1968).

\bibitem{PhysRev.166.1272}
R. A. Isaacson,
Phys. Rev. {\bf 166}, 1272 (1968).

\bibitem{Cai:2021jbi}
R.-G. Cai, X.-Y. Yang and L. Zhao,
Gen. Rel. Grav. {\bf 54}, 89 (2022).
[\arXiv[gr-qc]{2109.06864}]

\bibitem{Szabados_2009}
L. B. Szabados,
Living Rev. Relativ. {\bf 12}, 1 (2009).

\bibitem{Choptuik_1993}
M.~W.~Choptuik,
Phys.\ Rev.\ Lett.\  {\bf 70}, 9 (1993).

\bibitem{Gundlach_2007}
C.~Gundlach and J.~M.~Martin-Garcia,
Living Rev. Relativ.  {\bf 10}, 1 (2007).
[\arXiv[gr-qc]{0711.4620}]

\bibitem{Deppe:2018uye}
N. Deppe, L. E. Kidder, M. A. Scheel and S. A. Teukolsky,
Phys. Rev. D {\bf 99}, 024018 (2019).
[\arXiv[gr-qc]{1802.08682}]

\bibitem{Baumgarte:2019fai}
T. W. Baumgarte, C. Gundlach and D. Hilditch,
Phys. Rev. D  {\bf 123}, 171103 (2019).
[\arXiv[gr-qc]{1909.00850}]

\bibitem{Mendoza:2021nwq}
M. F. P. Mendoza and T. W. Baumgarte,
Phys. Rev. D {\bf 103}, 124048 (2021).
[\arXiv[gr-qc]{2104.03980}]

\bibitem{Zhang:2021nnn}
C.-Y.~Zhang, Q.~Chen, Y.~Liu, W.-K.~Luo, Y.~Tian and B.~Wang,
Phys. Rev. Lett. \textbf{128}, 161105 (2022).
[\arXiv[gr-qc]{2112.07455}]

\bibitem{Guo:2018yyt}
J.-Q.~Guo and H.~Zhang,
Eur. Phys. J. C \textbf{79}, 625 (2019).
[\arXiv[gr-qc]{1808.09826}]

\bibitem{He:2020ygx}
X.-K. He and N.-Q. Xie,
Class. Quant. Grav. \textbf{37}, 185016 (2020).
[\arXiv[gr-qc]{2005.04659}]

\bibitem{Afshar:2009pi}
M. M. Afshar,
Class. Quant. Grav. \textbf{26}, 225005 (2009).
[\arXiv[gr-qc]{0903.3982}]

\bibitem{Miller:2017udt}
W. A. Miller, S. Ray, M.-T. Wang and S.-T. Yau,
Class. Quant. Grav. \textbf{35}, 055007 (2018).
[\arXiv[gr-qc]{1708.07532}]

\bibitem{Gudapati:2020cqb}
N. Gudapati and S.-T. Yau,
Adv. Theor. Math. Phys. \textbf{25}, 101 (2021).
[\arXiv[gr-qc]{2009.04865}]

\bibitem{Liang}
C.-B. Liang and B. Zhou,
{\it Introduction to Differential Geometry and General Relativity}, Vol. 1 (Science Press, Beijing, China 2006).

\bibitem{Zhang:2016kzg}
C.-Y.~Zhang, Z.-Y.~Tang and B.~Wang,
Phys. Rev. D \textbf{94}, 104013 (2016).
[\arXiv[gr-qc]{1608.04836}]

\bibitem{Landau_Lifshitz}
L. D. Landau and E. M. Lifshitz,
{\it The Classical Theory of Fields}, Course of Theoretical Physics Series Vol. 2 (Pergamon Press, Oxford, UK 1971).

\bibitem{Poisson}
E. Poisson and C. Will,
{\it Gravity}, (Cambridge University Press, Cambridge, UK 2014).

\bibitem{Guo:2023cya}
J.-Q.~Guo, Y.~Hu, P.-P.~Wang and C.-G.~Shao,
Chin. Phys. C \textbf{48}, 065104 (2024).
[\arXiv[gr-qc]{2307.04372}]

\bibitem{Chris_1993}
D. Christodoulou,
Commun. Pure Appl. Math. \textbf{46}, 1131 (1993).

\bibitem{Price:1996sk}
R.~H.~Price and J.~Pullin,
Phys. Rev. D \textbf{54}, 3792 (1996).
[\arXiv{gr-qc/9601009}]

\bibitem{Choptuik_workshop_1993}
M. W. Choptuik,
\emph{Critical Behaviour in Scalar Field Collapse,}
in \emph{Proceedings of a NATO Advanced Research Workshop on Deterministic Chaos in General Relativity},
Springer Science+Business Media, LLC. Editors: D. Hobill and A. Burd and A. Coley, 155-175, 1993.

\bibitem{Guo:2020ked}
J.-Q.~Guo, L.~Zhang, Y.-W.~Chen, P.~S.~Joshi and H.~Zhang,
Eur. Phys. J. C \textbf{80}, 924 (2020).
[\arXiv[gr-qc]{2011.06792}]

\end{thebibliography}
\end{document}